\documentstyle[preprint,aps,eqsecnum]{revtex}   

\newcommand{\beq}{\begin{equation}} 
\newcommand{\eeq}{\end{equation}}
\newcommand{\beqa}{\begin{eqnarray}}
\newcommand{\eeqa}{\end{eqnarray}}

\def\sm{Standard Model}
\def\sign{\mbox{{\rm sign}}}

\def\Im{\mbox{{\rm Im}}}

\def\al{\alpha}
\def\be{\beta}

\def\plb#1{Phys.\ Lett.\ {\bf B #1}}
\def\prd#1{Phys.\ Rev.\ {\bf D #1}}
\def\prl#1{Phys.\ Rev.\ Lett. {\bf #1}}

\def\zpc#1{Z.~Phys.\ {\bf C #1}}

\def\ijmpa#1{Int.\ J.\ Mod.\ Phys.\ {\bf A #1}}

\begin{document}

\draft{\tighten
\preprint{
\vbox{
      \hbox{SLAC-PUB-7454}
      \hbox{hep-ph/9705356}}}

\bigskip
\bigskip

\renewcommand{\thefootnote}{\fnsymbol{footnote}}

\title{Removing Discrete Ambiguities in CP Asymmetry Measurements}
\footnotetext{Research supported
by the Department of Energy under contract DE-AC03-76SF00515}
\author{Yuval Grossman and Helen R. Quinn}
\address{ \vbox{\vskip 0.truecm}
Stanford Linear Accelerator Center \\
        Stanford University, Stanford, CA 94309 }

\maketitle

\begin{abstract}%
We discuss methods to resolve the ambiguities in CP violating phase angles
$\phi$ that are left when a measurement of $\sin 2 \phi$ is made. 
We show what knowledge of hadronic quantities will be needed to fully 
resolve all such ambiguities.
\end{abstract}

}
\newpage


\section{Introduction}

If we assume Standard Model unitarity there are two independent angles
in the ``unitarity triangle'', both of which are related to the 
underlying non-zero phases  of CKM matrix elements. 
We use the definition $\gamma = \pi - \beta - \alpha$,
where 
\beq
\alpha\equiv
\arg\left[-{V_{td}V_{tb}^*\over V_{ud}V_{ub}^*}\right], \qquad
\beta\equiv
\arg\left[-{V_{cd}V_{cb}^*\over V_{td}V_{tb}^*}\right],
\eeq
have simple interpretations as phases of particular 
combinations of CKM matrix elements.  

In $B$ factory experiments we seek to measure quantities that, in the
absence of physics from beyond the \sm, are simply related to
these angles.  Ignoring for the moment the effects of subleading
amplitudes, CP violating
asymmetries are proportional to $\sin2\phi$ where $\phi$ is
one of the angles of the triangle. In particular, the first two CP
asymmetries to be measured are likely to be in $B \to \psi K_S$ which
measures $\sin2\beta$, and in $B \to \pi^+\pi^-$ which measures
$\sin2\alpha$.  However, measurement of $\sin2\phi$ can only determine the
angle $\phi$ up to a four fold ambiguity: $\{\phi, \pi/2-\phi, \pi + \phi,
3 \pi/2 -\phi\}$ with the  
angles defined by convention to lie between $0$
and $2\pi$.  Thus, with two independent angles, there can be a priori a
total 16 fold ambiguity in their values as determined from CP asymmetry
measurements.  These ambiguities can limit our ability to test the
consistency between the measured value of these angles and the range
allowed by  other measurements interpreted in terms of the
\sm\ CKM matrix elements \cite{wolf}.
  
In any model where
the angles measured by the asymmetries in $B \to
\psi K_S$ and  $B \to \pi^+\pi^-$ are two angles of a triangle 
only 4 of the 16 choices are allowed, 
since the other combinations are incompatible
with this geometry \cite{nir-quinn}. Within the \sm, 
the present data on the CKM matrix
elements further reduce the allowed range,  implying that $2\beta$
is in the first quadrant ($0 < \beta < \pi/4$), that $0 <\alpha< \pi$, and
that there is a correlation between the values of $\al$ and $\be$ \cite{gn1}.
Thus, among the 16 possible solutions at most two, 
and probably only one, will
be found to be consistent with \sm\ results.

In the presence of physics beyond the \sm\ the values of the
``would be'' $\alpha$ and $\beta$ extracted from asymmetry measurements may
not fall within their \sm\ allowed range. Such new physics cannot be
detected if the values of the asymmetry angles happen to be related via the
ambiguities to values that do overlap the Standard Model range. Clearly, the
fewer ambiguous pairings that remain, the better our chance of recognizing
non-Standard Model physics should it occur.

One way to resolve these ambiguities is to measure asymmetries that depend
on very small angles \cite{nir-quinn,AKL}: 
$\arg[-{V_{cs}V_{cb}^*/V_{ts}V_{tb}^*}]$ or
$\arg[-{V_{cd}V_{cs}^*/V_{ud}V_{us}^*}]$.  
In this work we discuss other ways to resolve the ambiguities by measuring
asymmetries that relate to large angles 
only. That is not to say we discuss only easy
measurements. We will  later  briefly discuss the experimental
difficulties, but first we review the issue from a theoretical
perspective.  In addition to the {\it values} of $\sin 2\phi$,  
only the {\it signs} of $\cos 2 \phi$ and $\sin \phi$ for both
$\phi=\alpha$ and $\phi=\beta$ need to be determined.  These four signs
resolve the ambiguities completely:
\begin{itemize}
\item{%
$\sign(\cos 2 \phi)$ is used to resolve the $\phi \to \pi/2 - \phi$
ambiguity.  
}
\item{%
$\sign(\sin \phi)$ is used to resolve the $\phi \to \pi + \phi$
ambiguity.  
}
\end{itemize}

Several measurements which can determine $\sign(\cos 2 \phi)$ 
have been proposed \cite{nir-quinn,SnQu,wolf,Oliver}.  
Uncertainties in calculation of hadronic effects do not
affect the interpretations of these measurements,
although they do depend on the known value of hadronic
quantities such as the width and the mass of the $\rho$.
The determination of
$\sign(\sin\phi)$, however, cannot be achieved
without some theoretical input on hadronic physics.  Quantities that are
independent of hadronic effects  always appear as the ratio of a product of
CKM matrix elements to the complex conjugate of the same product. Such pure
phases are thus always twice the difference of phases of the CKM elements.
Any observable that directly involves a weak phase difference of two CKM
elements, $\phi$, (rather than $2 \phi$) also involves hadronic quantities
such as the ratio of magnitudes of matrix elements and 
the difference of their strong
phases. Thus, in order to determine the sign of $\sin\alpha$ or $\sin\beta$
some knowledge about hadronic physics is required.

We note that this is true even for our current knowledge of the \sm\
CP violating phase, $\sin \delta>0$ (where $\delta$  is the single
independent phase in the standard parametrization of the CKM matrix
\cite{pdg}).  In order to determine $\sign(\sin \delta)$ input on the sign of
$B_K$ is used \cite{nir-quinn}. The quantity $B_K$ is a ratio of hadronic
matrix elements. Its value is totally determined by the strong interactions
and thus, a-priori, is not reliably calculable. 
However, by now many methods of
determining $B_K$, including lattice calculations, all find that $B_K>0$,
though the range of allowed values is still quite large.  As a result, it is
now widely accepted that the sign of $B_K$ is reliable and thus that, in the
\sm, $\sin \delta > 0$.  

Many weak decay amplitudes include two terms with different weak phases. In
this work we show how the presence of a second term can 
be used to determine the
sign of $\sin\alpha$ and $\sin\beta$.  The needed theoretical input is the
sign of the real part of the ratio of the two amplitude terms
(excluding CKM elements).
The focus of this paper  is to examine what input assumptions are needed to
determine this sign, and discuss the status of these assumptions.  Our aim is
to clarify what is the minimum understanding of strong interaction effects
that will be needed to resolve the angle  ambiguities.  Our current arguments
alone cannot stand as a convincing reason to exclude an angle consistent with
the Standard Model range in favor of a choice that is not consistent.
However, were such a choice favored by this argument, it would at least pose
a serious challenge to theorists  to understand better the strong interaction
effects involved.  Eventually it may be that we have to piece together many
such puzzles to get a view of non-Standard Model physics from the low energy
frontier of $B$ decays.

In section 2 we review the general formalism of CP asymmetries in $B$
decays.  In section 3 we review methods to determine $\sign(\cos
2\phi)$.  In section 4 we explain how to determine $\sign(\sin \phi)$, and
what is the theoretical input that has to be supplied.  Finally, section 5
contains discussions and conclusions.


\section{General formalism}
In this section we present 
the general formalism of CP asymmetries in
$B$ decays. We start by explaining how we group
penguin and tree diagrams and then present the
needed formalism.

\subsection{Two-term weak decay amplitudes.}

The terms ``penguin'' and ``tree'' amplitudes are standard in the field for
weak decay amplitudes, but are actually only meaningful at the
short-distance, quark-diagram level. Our argument here
is quite  general and is
not in any way affected by the ambiguity inherent in these short distance
labels.  We group amplitude terms  together by weak phase, rather than by
individual diagrams. Then there is no need to attempt the unphysical
distinction between rescattering of a tree diagram and  a long-distance cut
of a penguin diagram. Further we use CKM unitarity to eliminate
one out of the  up, charm and top penguin diagrams terms.
In this way 
any $B$ decay amplitude, including all tree and penguin
diagrams, can be written as a sum of two terms, each with a definite weak
phase related to particular CKM-matrix elements.  The 
most convenient choice of how to group terms
depends on the final state quarks.  

For $b \to q \bar q s $ decays, for any final state $f$,
it is convenient to choose the two terms as
\beq \label{twoterms-s}
A_f^s = V_{cb}V_{cs}^* A_f^{ccs} + V_{ub}V_{us}^* A_f^{uus}.
\eeq
The second term here is Cabbibo suppressed compared to the first
and is negligible in most cases. 
For $b \to c \bar c s$ decays (e.g., $B \to \psi K_S$)
the  second term gets further suppression since the
dominant term includes a tree level diagram while the CKM-suppressed term 
contains only one loop (penguin) diagrams, namely, $A_f^{ccs} \gg A_f^{uus}$.
In $b \to u \bar u s$ decays the tree diagram contributes to the second term 
while the first term has  only penguin contributions and 
hence $A_f^{ccs} \ll A_f^{uus}$, thus  in this case there
is no clear hierarchy among the two terms.
 
For $b \to q \bar q d$  decays all the CKM coefficients
are of the same order of magnitude. It is then convenient to express
the amplitude as 
\beq\label{twoterms-d}
A_f^d = V_{qb}V_{qd}^* A_f^{qqd} + V_{tb}V_{td}^* A_f^{ttd},
\eeq
where $q=u$ or $c$ is chosen 
so that the  first term includes any tree diagram
contribution for the channel in question. (When there is no tree
diagram the choice is arbitrary.)
The second term  here has a weak phase predicted
in the \sm\ to be half the weak phase of the
mixing amplitude. Thus, only one unknown weak phase
difference enters the analysis when the amplitude is written in this way.

For any given channel at most one of these two  terms has a tree diagram
contribution.  The tree diagram is generally expected to be the
dominant contribution to any $A_f^{qqq'}$ for which it is non-zero,
so we will call this the ``tree-dominated'' term to remind the
reader that it also contains a  difference of  loop (or penguin)
contributions with the same weak phase.  We then refer to the other
term, which has no tree diagram contribution,  as the ``penguin-only'' term.

We note, as an aside,  that the two-term structure of decay amplitudes
can also accommodate any beyond-Standard-Model physics contribution, 
since any additional term in a decay amplitude, whatever its phase, 
can always be written as a sum of two terms of definite phase with 
(possibly negative) real magnitudes.  
The difference between \sm\ physics  and non-Standard-Model
physics then comes down to  the expected relative sizes of the two terms.
These expected  sizes  are, in general, dependent on our understanding 
of hadronic matrix elements. This just shows once again how difficult it 
could be to  recognize the presence of  non-\sm\ physics.  
The only reliable way to find new effect in decay amplitudes is
to examine cases in which  a single term significantly dominates 
the weak decay amplitude in the \sm\ \cite{gw}.

\subsection{General formalism}
 
Here we recall the general formalism of CP asymmetries in
$B$ decays. We use the standard 
notations \cite{Yossi}. We assume the Standard Model all the way.

The time dependent CP asymmetry in $B$ decays
into a final CP eigenstate state $f$ is defined as \cite{Yossi}
\beq
a_f(t) \equiv 
\frac{\Gamma[B^0(t) \rightarrow f] - \Gamma[\bar B^0(t) \rightarrow f]}
     {\Gamma[B^0(t) \rightarrow f] + \Gamma[\bar B^0(t) \rightarrow f]},
\eeq
and is given by
\beq \label{acp}
a_f(t) = a_f^{\cos} \cos(\Delta Mt) + a_f^{\sin} \sin(\Delta Mt),
\eeq
with
\beq
a_f^{\cos} \equiv {1-|\lambda|^2 \over 1 + |\lambda|^2}, \qquad
a_f^{\sin} \equiv {- 2 \,\Im \lambda \over 1 + |\lambda|^2},\qquad
\lambda \equiv {q \over p} {\bar A \over A},
\eeq
where $p$ and $q$ are the components of the interaction
eigenstates in the mass eigenstates,
$|B_{L,H}\rangle=p |B^0 \rangle \pm q |\bar B^0 \rangle$, and
$A(\bar A$) is the $B_d(\bar B_d) \to f$ transition amplitude \cite{Yossi}.
The time-dependent measurement can  separately determine 
$a_f^{\cos}$ and $a_f^{\sin}$.
We always consider decays with a leading tree diagram amplitude.
Then, we write the amplitude as
\beq
A=A_T e^{i\phi'_T} e^{i\delta_T} + A_P e^{i\phi'_P} e^{i\delta_P}, \qquad
\bar A=A_T e^{-i\phi'_T} e^{i\delta_T} + A_P e^{-i\phi'_P} e^{i\delta_P}
\eeq
where $T$ and $P$ stand for the tree-dominated and penguin-only  
terms respectively. 
The weak phases of the decay amplitudes, $\phi'_T$ and $\phi'_P$ 
are convention dependent, as is  $\arg(q/p)$ but the differences
$\phi_T = \phi'_T - \arg(q/p)/2$ and 
$\phi_P = \phi'_P - \arg(q/p)/2$ are
convention independent quantities that we seek to determine.
Similarly, the strong phases are all subject to arbitrary
redefinitions, only the relative strong phase of the two terms 
$\delta \equiv \delta_P - \delta_T$ is a
physically meaningful quantity. We have introduced strong phases for
each term so that we can always fix both $A_T $ and $A_P$ to be real
quantities, independent of any phase convention choice.
We then define the real quantity 
\beq \label{rdef}
r \equiv {A_P \over A_T}.
\eeq
Note that we allow $r<0$.
The CP violation sensitive quantity $\lambda$ is then 
\beq \label{reslam}
\lambda = 
\eta_f {e^{-i\phi_T} + r e^{-i\phi_P} e^{i\delta} \over
 e^{i\phi_T} + r e^{i\phi_P} e^{i\delta}}. 
\eeq
Here $\eta_f$ is the CP parity of the final state. In particular,
$\eta_{\psi K_s}=-1$ and $\eta_{\pi^+\pi^-}=\eta_{D^+D^-}=1$.

For $b \to c \bar c s$ decays,  leading for example 
to the final state $\psi K_S$, the penguin-only term 
is Cabbibo suppressed and can be safely neglected. 
Thus $r=0$ should be an excellent approximation 
and we get the well known result
\cite{Yossi}
\beq \label{arz}
a_f^{\cos}=0, \qquad a_f^{\sin} = \eta_f \sin 2 \phi_T.
\eeq
We next consider $b \to q \bar q d$ decays, leading for example 
to the final states $B \to \pi^+ \pi^-$ or 
$B \to D^+D^-$. Here, by definition, $\phi_P = 0$
since the penguin contributions with a different weak phase are subsumed in
$A_T$. Then 
\beq \label{apz}
a_f^{\cos} =
{2r\sin\phi_T\sin\delta \over 1+r^2+2r\cos\phi_T\cos\delta}, \qquad
a_f^{\sin} = \eta_f {\sin 2\phi_T + 2r\sin\phi_T\cos\delta \over 
1+r^2+2r\cos\phi_T\cos\delta}.
\eeq

\section{Determining $\sign(\cos 2\phi)$}
In this section we review measurements that can be used to extract
$\sign(\cos 2\alpha)$ and $\sign(\cos 2\beta)$. 
These signs resolve the $\phi \to \pi/2 -\phi$ ambiguities.

\subsection{$B \to \rho \pi$}

All the three decays $B \to \rho^+ \pi^-$, $B \to \rho^- \pi^+$ and $B \to
\rho^0 \pi^0$ can lead to a $\pi^+\pi^-\pi^0$ final state.  Due to
interferences between these channels sufficient information is encoded in
the $B \to \rho \pi$ decays to distinguish between the $\alpha$ and $\pi/2
-\alpha$ choices.  This was shown in Ref. \cite{SnQu}, where it was
explained how both $\sin 2 \alpha$ and $\cos 2 \alpha$ can be measured using
a full Dalitz plot distribution analysis. To resolve the ambiguity one needs
only to fix the sign of $\cos 2 \alpha$, which should be relatively easy to
achieve.

We do not repeat here the detailed explanations of Ref. \cite{SnQu}.  In
that work it was shown that there are several observables that, in the
absence of penguins, directly measure $\cos 2\al$. (These observables all
involve the imaginary part of an overlap between two different Breit-Wigner
functions describing two different charges of $\rho$ meson.) The presence of
penguins spoils the simple relationship between these quantities and   $\cos
2\al$. However, even when penguin terms are present, there is enough
information in the interference regions to determine the sign of  $\cos
2\al$. A multiparameter fit can obtain a preferred choice between $\alpha$
and $\pi/2 - \alpha$, even allowing for arbitrarily large penguin
contributions.

Here, and throughout this paper, we neglect the effects of electroweak
penguins. These give a correction to isospin-based treatments for isolating
certain CKM factors. The isospin structure of the amplitudes contributing
to $\rho \pi$ decays is used to isolate terms with isospin two, because they
receive no contribution from QCD penguin graphs, and
hence show pure $\sin 2\alpha$ and/or  
$\cos 2 \alpha$ dependence. Electroweak penguin
graphs can give isospin two parts but the relevant contributions here are
expected to be quite small and hence unlikely to confuse the extraction
of the sign of $\cos 2 \alpha$.

Experimentally, the $\cos 2 \al$  
determination involves fitting
parameters to the contributions of a broad resonance. Under
these resonances there are non-resonant $B$ decay contributions which
must also be fit in order to extract the relevant resonant
effects.  The question of how best to parameterize these non-resonant
contributions is under study \cite{babarphys}. 
It will have to be resolved to extract
useful results from these channels.

\subsection{$B \to D D^{**}$}

The idea of using overlapping decays to add information on $\cos2\phi$ can
be in principle applied to $B$ decays to higher $D$ resonances
\cite{Oliver}.  In that case, a full Dalitz plot distribution of  
$D^{(*)}D^{(*)}\pi$
final states can be used to determine the sign of $\cos 2\be$.  Since the
$D^*$ are rather narrow the interference effects are probably too small to
be detected in $B \to DD^*$ since there is essentially no overlap kinematic
region between different $D^*$'s.  The $B \to D^{(*)} D^{**}$ 
decays are better candidates. 
The $D^{**}$ widths are larger and the effect may be 
measurable.  More details are expected to be given in Ref. \cite{Oliver}.
Once again, it may be a problem to parameterize non-resonant 
$D^{(*)}D^{(*)}\pi$ that 
contribute in the same region as the resonances and could potentially
destroy the analysis.

\subsection{$B^\pm \to DK^\pm$}
The angle $\gamma$ satisfies the condition
\beq
\al+\be+\gamma=\pi\ (\mbox{\rm mod}\ 2\pi).
\eeq
Since $\gamma$ is defined modulo $2\pi$,
the 16 possibilities for $\alpha$ and $\beta$ 
result in an eightfold ambiguity in $\gamma$.
These eight values give two different values for
$\cos  2\gamma$ and four different values for $\sin  2\gamma$.
Thus, by measuring $\cos  2\gamma$ or $\sin  2\gamma$ some of the 
ambiguities can be resolved.
Here we focus on $\cos  2\gamma$ and in the next subsection
we discuss $\sin  2\gamma$.

The value of $\cos 2\gamma$ can be used to resolve
some combination of the $\phi \to \pi/2-\phi$ ambiguities.
The trigonometric identity 
\beq \label{signgam}
\cos 2\gamma= \cos 2\beta \cos 2 \al-\sin 2 \al \sin 2 \beta,
\eeq
implies that the transformations $\beta \to \pi/2-\beta$
or $\alpha \to \pi/2-\alpha$ (but not both)
change the value of $\cos 2\gamma$. 
As we assume that $\sin 2 \beta$ and $\sin 2 \al$ are known, 
$\cos2\gamma$ can distinguish between the two cases
$\{\al,\be\}, \{\pi/2-\al,\pi/2-\be\}$ or 
$\{\pi/2-\al,\be\}, \{\al,\pi/2-\be\}$.
Thus, for example, if $\cos 2 \al$ in known from the 
$B \to \rho \pi$ analysis,
the sign of $\cos 2\beta$ can be determined from the measurement 
of $\cos 2\gamma$.

Several methods to extract $\sin^2 \gamma$ (or equivalently $\cos 2\gamma$)
using $B^\pm \to DK^\pm$ decays \cite{DKGrWy,DKDu} or $B_s$ decays
\cite{moreBs} have been proposed \cite{BuFl}. For the purpose of
illustration,  below we concentrate on the method of \cite{DKGrWy}.  This
method uses measurements of six $B^\pm \to DK^\pm$ decay rates to extract
$\cos 2\gamma$ up to a two fold ambiguity.  This two-fold ambiguity is due
to an unknown strong phase. In general, this ambiguity can be removed by
applying the same analysis for  several final states \cite{DKGrWy}
with the same flavor quantum numbers as $DK^\pm$. All these modes have the 
same weak phase but, in
general, different strong phases. Thus, only one solution of $\cos 2\gamma$
is consistent in all the modes while the second (incorrect) one should be
different in the different modes, since strong phases differ from one
mode to another.  

We note that even if we have a two-fold ambiguity in $\cos 2\gamma$ 
because we have studied only a single final state system, 
the incorrect value
of $\cos 2\gamma$ should not be the same as that obtained using the
incorrect value of $\beta$ or $\alpha$. 
In that case there are going to be two possible
solutions for $\cos 2\gamma$ from the $B^\pm \to DK^\pm$ measurement, and
two predictions arising from the measurements of $\sin 2 \beta$
and $\sin 2 \al$.  In
general, only one of the solutions will coincide and the other not. 
Choosing the one that coincides is sufficient to resolve the ambiguity
in the $\cos 2\gamma$ measurement and at the same time to fix the relative
sign of $\cos 2\al$ and $\cos 2 \be$.

\subsection{$B_s \to \rho K_S$}

The time dependent CP asymmetry in certain $B_s$ decays  (e.g., $B_s \to
\rho K_S$) directly measures $\sin 2 \gamma$ if the penguin-only term
in the decay amplitude is neglected. 
A measurement of $\sin 2\gamma$
would  determine the signs of $\cos 2\beta$ and 
$\cos 2\alpha$ \cite{nir-quinn}, assuming their magnitudes are known.  
The trigonometric identity 
\beq \label{signgamco}
\sin 2\gamma= -(\cos 2\beta \sin 2\al+\cos 2 \al \sin 2 \beta),
\eeq
implies that either or both
of the transformations $\beta \to \pi/2-\beta$ and $\alpha \to
\pi/2-\alpha$, change the value of $\sin
2\gamma$.  Thus, the signs of both
$\cos 2 \al$  and $\cos 2 \be$  can be determined, once $\sin 2\gamma$ is
known.

Experimentally, it will be very hard, if at all possible, 
to measure this asymmetry. 
In addition, the penguin-only term is expected to 
be significant in $b \to u\bar ud$
decays, making the relationship between the asymmetry and the angle $\gamma$
more complicated \cite{BuFl}. These problems imply that the
methods we mentioned before are better than
the time dependent CP asymmetry in
$B_s \to \rho K_S$ decay for determining $\gamma$ \cite{BuFl}. 
However, all these other methods determine $\cos 2 \gamma$. 
The justification to study the time dependent CP asymmetry in
$B_s \to \rho K_S$ is that it probes a different functional dependence of 
$\gamma$, namely, $\sin 2 \gamma$.

As we need only to choose between few discrete choices of $\gamma$
the problems mentioned before may not be so severe in our case.
By the time measurement of the CP asymmetry in
$B_s \to \rho K_S$ is feasible we will probably already know  the rough value
of the penguin contribution, from its relationship to similar effects in $B
\to \pi\pi$, extracted via isospin analysis, and those determined from fits
to  $B_d \to \rho \pi$. If $\cos 2 \gamma$ is already measured as discussed
above, then we need this measurement 
only to distinguish between the two values of the sign of 
$\sin 2 \gamma$. 
In general only one sign will be consistent with the allowed range
for the ratio of penguin-only to tree-dominated terms, so the ambiguity will
be resolved even though an a-priori measurement of $\sin 2 \gamma$ cannot 
be achieved.

\section{Determining $\sign(\sin \phi)$}

In this section we discuss how $\sign(\sin \alpha)$ and $\sign(\sin \beta)$
can be determined. These signs resolve the $\phi \to \pi + \phi$ ambiguity.
As we already explained, this ambiguity cannot be resolved in any 
theoretically
clean way. Some knowledge of hadronic physics is  always needed. In the
following we describe several methods that can be used to resolve the
ambiguity, and explain what is the needed theoretical input.

In order to get sensitivity to $\sign(\sin \phi)$ we focus on cases where
two terms with different weak phases are involved in the decay amplitude. 
Then, in principle, the relative phase between these two terms can be
determined. However, there is also a relative strong phase between these two
terms. Therefore, theoretical input is required in order to disentangle the
strong and the weak phases. The relevant hadronic quantity is found to be
the sign of $r\cos\delta$, that is the sign of the  real part of the ratio
of the two amplitude terms (excluding weak phases).

\subsection{$B \to \psi K_S$ vs $B \to D^+D^-$}

In the case of the angle $\beta$ we have one class of measurements, from 
$b \rightarrow c \bar c s$  processes such as $B \to \psi K_S$, that have
very small $r$. For these channels Eq. (\ref{arz}) with $\phi_T=\beta$ 
is valid  and the asymmetry measurement determines $\beta$ up
to the usual four-fold ambiguity \cite{Yossi}
\beq \label{aPSIKS}
a_{\psi K_S}^{\sin} = -\sin 2 \beta.
\eeq

The other class of measurements is from $b \to c \bar c d$ decays such as
$B \to D^+ D^-$.
In this case we expect $r$ to be significantly larger 
and Eq. (\ref{apz}) with $\phi_T=\beta$ is valid. For simplicity we
will here give results valid only to leading order in $r$, however 
we have checked that the full
expression contains enough information to avoid this approximation if
needed. We get
\beq \label{aDD}
a_{D^+D^-}^{\sin} =
\sin 2\beta - 2r_{DD}\cos 2\beta \sin\beta\cos\delta_{DD}.
\eeq
where $\delta_{DD}$ is the strong phase difference between the
tree-dominated and penguin-only $B \to D^+D^-$
amplitudes, and $r_{DD}$ 
is the signed ratio of their magnitudes.
Comparing Eqs. (\ref{aDD}) and (\ref{aPSIKS}) we find
\beq \label{signfi}
a_{\psi K_S}^{\sin} + a_{D^+D^-}^{\sin} =
-2 r_{DD}\cos\delta_{DD} (\cos 2\beta \sin\beta).
\eeq
It is clear from this expression that we can fix the sign of
$\sin\beta$ only if we know  the sign of $\cos 2 \beta$ and, in
addition, the sign of $r_{DD}\cos\delta_{DD}$. We assume the first of these
is given by the methods discussed in the previous section.

Currently, there is no reliable way to determine the sign of the 
real part of the ratio
of hadronic matrix elements $(r_{DD}\cos\delta_{DD})$. 
In order to proceed, we assume factorization. 
(We will discuss the reliability of this and 
subsequent assumptions later.)
Assuming factorization and that the top 
penguin is dominant, we can infer from the results of
Ref. \cite{KP}, $r_{DD} < 0$. 
Within the factorization approximation the relevant strong phases (almost)
vanish, so that $\delta_{DD}\simeq 0$, and hence the sign of 
$r_{DD}\cos\delta_{DD}$ is given by the sign of $r_{DD}$. 

Assuming $r_{DD}\cos\delta_{DD}<0$
as given by the factorization calculation we get
\beq \label{signbeta}
\sign(a_{\psi K_S}^{\sin} + a_{D^+D^-}^{\sin}) = 
\sign (\cos 2\beta \sin\beta).
\eeq
Note, in particular, that the \sm\ predicts
$\cos 2\beta \sin\beta > 0$, and therefore also that the asymmetry in
$D^+D^-$ is smaller in magnitude than the asymmetry in $\psi K_S$ 
(and opposite in sign). 

We need only measure the sign of the sum of the two asymmetries 
to resolve the ambiguity. Even this may not be
an easy task if $r_{DD}$ is small,  
however a recent estimate found that in the \sm\
$3\% \lesssim r_{DD} \lesssim 30\%$ \cite{CFMMS}, 
and certainly in the upper end of
this range the required sign should be measurable.


\subsection{$B \to \rho \pi$ vs $B \to \pi^+\pi^-$}
We first explain how to
get $\sin 2\alpha$ uniquely out of the $B \to \rho \pi$ decays 
without uncertainties due to penguin only terms.
Then, the comparison with the asymmetry in  $B \to \pi^+\pi^-$
can be used to determined $\sign(\sin\al)$ using
a similar approach to that discussed for $\beta$ above. 

While the experiment may well
proceed to determine all the various amplitudes and phases simultaneously by
a maximum likelihood fit, it is instructive to inspect the expressions
analytically to see what combination of terms actually enters into the 
measurement of $\sin 2\alpha$. We follow the treatment of \cite{SnQu}
and write 
\beqa
A_3=A(B^0 \to \rho^+ \pi^-)=T_3+P_1+P_0\,, \quad && \quad
\bar A_3=A(\bar B^0 \to \rho^- \pi^+)=\bar T_3+\bar P_1+\bar P_0\,, 
\\
A_4=A(B^0 \to \rho^- \pi^+)=T_4-P_1+P_0\,,  \quad && \quad
\bar A_4=A(\bar B^0 \to \rho^+ \pi^-)=\bar T_4- \bar P_1+\bar P_0\,,
\nonumber \\
A_5=A(B^0 \to \rho^0 \pi^0)=T_5-P_0\,,  \quad && \quad
\bar A_5=A(\bar B^0 \to \rho^0 \pi^0)=\bar T_5-\bar P_0\,, 
\nonumber 
\eeqa
where $T_i$ is the tree-dominated amplitude and $P_1$ and $P_0$ are the 
(suitably rescaled) penguin-only contribution for isospin one and isospin 
zero respectively. The CP conjugate amplitudes
$\bar A_i$, $\bar T_i$ and $\bar P_i$ differ from 
the original amplitudes, $A_i$, $T_i$ and $P_i$  
only in the sign of the weak phase of each term.
We further define
\beqa
A_{sum}&\equiv& A_3+A_4+2A_5=\left(|T_3|e^{i\delta_3}
+|T_4|e^{i\delta_4}+2|T_5|e^{i\delta_5}\right)e^{i\phi'_T}, \\
\bar A_{sum}&\equiv& \bar A_3+\bar A_4+2\bar A_5=\left(|T_3|e^{i\delta_3}
+|T_4|e^{i\delta_4}+2|T_5|e^{i\delta_5}\right)e^{-i\phi'_T}. \nonumber 
\eeqa
Here, $\delta_i$ is the strong
phase of $T_i$, and $\phi'_T$ is the common weak phase of the 
tree-dominated terms. 
We see that $\bar A_{sum} = A_{sum}e^{-2i\phi'_T}$.
From Table I of Ref. \cite{SnQu} we see that both
$A_{sum} A_{sum}^*$ and $\Im\left(q \bar A_{sum}\, p^* A_{sum}^*\right)$
are observables. (Note that $q$ as defined in Ref. \cite{SnQu} is equal
to $\sqrt{2} q p^*$ in our standard notation.)
In particular, we see that from the data we can extract
\beq \label{aRHOPI}
a^{Dalitz}_{\rho\pi} \equiv
-\Im \left({q \over p} {\bar A_{sum} \over A_{sum}}\right) = -\sin 2\al,
\eeq
where for the last equation we used $|q/p|=1$ and $\phi_T=\pi-\al$.
Eq. (\ref{aRHOPI}) shows that 
$\sin 2\alpha$ can be extracted using $B \to \rho \pi$ decays 
without penguin pollution.
We emphasize that in order to obtain this result we did not have to assume
that the top penguin is dominant. All penguin terms are included, either
as a subdominant part in the tree-dominated amplitudes, or in the 
penguin-only term.

Alternately, $B \to \pi\pi$ decay modes can also be used to extract $\sin 2
\al$ without hadronic uncertainties using isospin analysis.  The needed
measurements are the time-dependent rate for  $B \to \pi^+ \pi^-$ together
with the time-integrated rates of  $B^0 \to \pi^0 \pi^0$, $B^+ \to \pi^+
\pi^-$ and their conjugate decays \cite{GrLo}, and a geometrical construction
then allows extraction of $\sin2 \alpha$.  However, discrete ambiguities in
this construction imply that $\sin 2 \al$ can only be extracted up to certain
discrete choices, which correspond also  to differences in the relative phase
and the ratio of magnitudes of certain tree-dominated and  penguin-only 
terms (but not the
same combinations as we identify below).  The determination from $\rho \pi$
does not suffer from this problem.  (These ambiguities could in principle be
removed by a precise measurement of the time dependent asymmetry in $B\to
\pi^0\pi^0$ \cite{GrLo}, but this measurement is unlikely.)

Now, assuming we have determined $\sin2 \alpha$, 
we look again at the $B \to \pi^+ \pi^-$
decay, here using the interference of the two terms in the amplitude to
determine the sign of $\sin\alpha$, just as we did in the $D^+D^-$
case for $\be$. Here,  $\phi_T = \pi-\alpha$ and $\phi_P = 0$,
and Eq. (\ref{apz}) gives the asymmetry. Once again, for simplicity, 
we work to leading
order in $r$, but this approximation can be avoided if
needed. We get
\beq \label{aPIPI}
a_{\pi^+\pi^-}^{\sin} = 
-\sin 2\al - 2r_{\pi\pi}\cos 2\al \sin\al \cos\delta_{\pi\pi},
\eeq
where $\delta_{\pi\pi}$ is the strong phase difference between the
tree-dominated and penguin-only $B \to \pi^+\pi^-$
amplitudes, and $r_{\pi\pi}$ 
is the signed ratio of their magnitudes.
Comparing Eqs. (\ref{aRHOPI}) and (\ref{aPIPI}) we get
\beq \label{signfial}
a_{\pi^+\pi^-}^{\sin} - a^{Dalitz}_{\rho\pi} =
-2 r_{\pi\pi}\cos\delta_{\pi\pi} (\cos 2\al \sin\al)
\eeq
Thus, once $\sign(r_{\pi\pi}\cos\delta_{\pi\pi})$ is known, 
the measurements will
determine $\sign(\cos 2\alpha\sin\alpha)$. 
If the $\sign(\cos 2\al)$ is known from the treatments discussed above, 
$\sign(\sin\al)$ is then
determined; if not, at least the fourfold ambiguity of 
$\{\sign(\cos2\alpha), \sign(\sin\alpha)\}$ is reduced
to a two-fold ambiguity.

Again, there is as yet  no reliable way to calculate the sign of 
$r_{\pi\pi}\cos\delta_{\pi\pi}$. 
Therefore, we turn to the short-distance calculation with factorization  
to determine \cite{KP} that
$r_{\pi\pi} < 0$ and that $\delta_{\pi\pi}$ is very small.
This then gives 
\beq \label{signal}
\sign(a_{\pi^+\pi^-}^{\sin} - a^{Dalitz}_{\rho\pi}) =
\sign(\cos 2\al \sin\al).
\eeq
With the knowledge of $\cos 2 \alpha$
this difference can be used to fix the sign of $\sin \al$.


\subsection{CP asymmetries in inclusive decays}

In the above, the main obstacle in getting theoretically clean predictions
is that we do not have a reliable way to calculate the ratio of the relevant
hadronic matrix elements. An alternative way, which does not suffer
from this problem, is to measure asymmetries in semi-inclusive
decays, e.g. to all states with a given flavor content \cite{bbd}. 
Here matrix elements are not needed. However a crucial
assumption in this case is that the semi-inclusive 
measurements are described by
the quark level calculations, which are needed to determine  $\xi$:
the fraction of CP-odd  final states. The quantity $1-2\xi$ is
referred to as the ``dilution factor''.  The assumption, 
called  local quark-hadron duality, that the 
quark-diagram kinematics are unaltered by hadronization,
is essential to this calculation and is not well
justified.  In addition, we are convinced that
full semi-inclusive measurements are not experimentally feasible,
some data cuts will be needed. The effect of such cuts on the ratio of 
CP-even to CP-odd contributions is difficult to calculate and likely to be
even more sensitively dependent on the local quark-hadron  
duality assumption.

However, our game here is to determine signs, so we can possibly use
these methods despite large uncertainties in the calculation of the
relevant dilution factors, as long as the sign of $(1-2\xi)$ is
reliably determined. The hope
is that by the time the inclusive measurements will be carried out, we will
have  consistency checks that will either support or rule out local duality.
For example, the inclusive asymmetry calculations are similar
to that of the $B_s$ width difference \cite{bbdGamBs}.
If future 
measurements of the $B_s$ width difference agree with this calculation, 
it would support the local duality assumption.

A potentially useful measurement is the asymmetry in the
$B_d \to D X$ where $X$ is multi pion state with no $K$ meson contributions.
Such  decays are governed by the
$b \to c \bar u d$ and $b \to u \bar c d$ transitions.
The inclusive calculation gives \cite{bbd}
\beq
a_{c \bar u d \bar d}^{\sin} = -(1-2\xi) 
\left|V_{cd}V_{ub} \over V_{ud}V_{cb}\right| \, \sin(\al-\be).
\eeq
On the practical side,  we note that  the large inclusive rate may 
help compensate the CKM suppression of the asymmetry.
We see that the $\al \to \pi+\al$ or $\be \to \pi+\be$ transformations
(but not both) will change the sign of the result. 
The quantity $(1-2\xi)$ is calculated to be about 0.21 \cite{bbd},
but the range of
uncertainty on this quantity, and its dependence on the necessary
experimental cuts remains to be explored. If we can convince ourselves
that we know the sign of this quantity, as calculated for the specific
data sample used to determine the asymmetry, we can use such a measurement to
reduce the set of ambiguous choices for the two angles. 
Perhaps one way to proceed will be to explore, both in the theory and
in the data, the sensitivity of the signs to changes in the selected sample.

Another measurement that can be useful is that of
$B_s$ decays governed by the
$b \to c \bar u s$ and $b \to u \bar c s$ transitions.
For this case Ref. \cite{bbd} found
\beq
a_{c \bar u s \bar s}^{\sin} \approx (1-2\xi)
\left|V_{cs}V_{ub} \over V_{us}V_{cb}\right| \, \sin(\al+\be),
\eeq
where here $1-2\xi\approx 0.28$ \cite{bbd}.
Again, the $\al \to \pi+\al$ or $\be \to \pi+\be$ transformations 
(but not both)
will change the sign of the result. 
Note that unlike the previous case, here the CKM suppression
is not very small.
However, asymmetries in $B_s$ decays are expected to be harder to measure.
Once again the dilution factor calculation 
needs to be further explored to determine whether
the sign of this quantity can reliably be calculated.


\subsection{Remarks about the theoretical assumptions}

We here examine the points at which it is important to clarify our
theoretical understanding if we are to use the results of $B$ factory
experiments to look for indications of non-\sm\ physics.  Our arguments can
be strengthened by a combination of improved calculational methods (such as
lattice calculations of matrix elements) and by testing the implications of
similar arguments in a variety of channels, in addition to those studied for
the CP studies. It is to be hoped that, by the time we have sufficient data
to perform the measurements described above, both of these avenues will have
been explored and our arguments, e.g. on the sign of
the $r\cos\delta$ terms, either discredited or more firmly
established. The point of this paper is that we need to pursue this further
understanding to resolve the ambiguous choices.

We will discuss here the exclusive final states. 
There, we use factorization to calculate
the sign of $r\cos\delta$.  Here we discuss why it is plausible that 
this sign is correctly predicted by the factorization calculation.
Our calculation uses the operator product expansion
approach, which is rigorous, but adds to it the less rigorous ingredients of
a model to calculate matrix elements. We apply this model only in 
color-allowed decays where the outcome is insensitive to the variation of the
parameter governing the relative contribution of color-suppressed terms.

The factorization approximation treats each quark-antiquark combination
separately, the only  strong phase, in this approximation, is a small
effect that arises from cuts of the short-distance penguin diagrams 
involving $u$ or $c$ quarks. Thus, $\delta \approx 0$.
To go beyond the factorization approximation we consider a two step
picture in which the decay and hadronization occurs as  calculated in
the factorization approximation but (elastic and inelastic)
final state rescattering are allowed. 
While here we present only the $D^+D^-$ final state, similar treatment 
apply  also to the $\pi^+\pi^-$ final state with similar conclusions.
The way to proceed
is to work in the  isospin basis. Each of the terms $A_T e^{i\delta_P}$
and $A_P e^{i\delta_P}$
has two  isospin contributions (labeled by 
the final state isospin $I_f = 0,1$). 
These terms acquire strong phases through rescattering effects.  
We emphasize that the rescattering phases for 
the same isospin channel can be different in the penguin-only and 
tree-dominated terms. 
These amplitudes have different overlap between the 
$D^+D^-$ state and the other
hadronic states with the same charm-quark content and isospin.
Because the light quark content in $D^+D^-$
is $d \bar d$ we know that in both the tree-dominated and
penguin-only terms separately the two isospin contributions are 
equal in magnitude. 
Thus, the effect of rescattering can be taken into account by 
writing the tree-dominated and penguin-only  amplitudes as
\beq
A_Te^{i\delta_T} \cos\delta^{01}_T, \qquad
A_Pe^{i\delta_P} \cos\delta^{01}_P.
\eeq
Here the phases are given by 
\beq
\delta_X=(\delta^0_{X} + \delta^1_{X})/2, \qquad
\delta^{01}_X = (\delta^0_{X} - \delta^1_{X})/2,
\eeq
where $\delta^i_{X}$ is the phase
shift of the isospin $i$ term in the $X=T,P$ amplitude. 
Thus, after rescattering, we find  
\beq \label{shifts}
r_{DD} = r_{DD}^{fact}
{\cos\delta^{01}_P\over\cos\delta^{01}_T}, \qquad
\cos\delta_{DD}=\cos(\delta_T-\delta_P),
\eeq
where $r_{DD}^{fact}$ is $r_{DD}$ as calculated using
factorization.
Thus, the sign of $r_{DD}\cos\delta_{DD}$ is unchanged by rescattering 
if the relevant phase shifts are all
sufficiently small that the cosines in Eqs. (\ref{shifts}) are all positive.

It seems to be a reasonable assumption that all the relevant strong 
phases are small. There are no known
nearby resonances with isospin 0 or 1 in the spin zero partial wave in
the $D^+D^-$ system at the $B$ mass. Furthermore, some cross checks on this
argument are
available. The rates of $D^+ D^-$ and $D^0 \bar D^0$
productions are given by
\beqa
\Gamma(B \to D^+ D^-) &=& 
\left|A_T\cos\delta^{01}_Te^{i\delta_T}e^{i\phi'_T} +
A_P\cos\delta^{01}_Pe^{i\delta_P}e^{i\phi'_P}\right|^2. \\
\Gamma(B \to D^0 \bar D^0) &=& 
\left|A_T\sin\delta^{01}_Te^{i\delta_T}e^{i\phi'_T} +
A_P\sin\delta^{01}_Pe^{i\delta_P}e^{i\phi'_P}\right|^2. \nonumber 
\eeqa
If the $D^0 \bar D^0$ rate is small compared to the $D^+D^-$ rate it provides
some confirmation that the rescattering phases $\delta^{01}_T$ and
$\delta^{01}_P$ are small. 

Direct CP violation effects in these channels depend on the same 
rescattering phases
and can be predicted in terms of the same parameterization. Such
effects are proportional to $\sin\delta$ and so are small if all
rescattering effects  are small. Large direct CP violations in the
$D^+D^-$ or $\pi^+ \pi^-$ channels would be a reason to mistrust our
argument for the sign of $r\cos\delta$. However, 
small direct CP violations are
consistent with, but not a convincing argument for small $\delta$.
An interesting example would be if
$\sin\delta$ is found to be small in several channels with the same 
quark content
(e.g. $D D$, $D D^*$ and $D^* D^*$).
Then, we would have to  conclude that
either $\delta \sim 0$  or $\delta \sim \pi$ in each of these channels. 
There is no reason to believe that any 
rescattering strong phases should
be close to $\pi$ and it is even
less likely that several at once have this value. 
However, due to the arguments for factorization,
it is quite plausible that all of them are small at the same time.

To conclude: the needed theoretical input is the sign of
$r\cos\delta$. Here, we argue that it is plausible that
the correct sign can be predicted by factorization in color-allowed 
channels. Moreover, some 
cross-check can be done.
However, we emphasize again that we believe that 
there is currently no reliable way to determine this sign.

\section{Final remarks and conclusions}

Our goal is to find physics beyond the \sm.  While in this paper we present
our results as a way to resolve the discrete ambiguities 
in the values of $\alpha$ and $\beta$, it
should be remembered that in the context of the 
\sm, because of constraints from other measurements,
there is only two fold ambiguity in
$\alpha$ and no ambiguity in $\beta$. The importance of resolving the
ambiguities is to expose a possible inconsistency with the \sm\ values. 
This will then indicate new physics.

When looking for new physics, one should try to assume as little as possible
about its nature. Here, we allowed any kind of new physics. 
This new physics can be (any combination of) new contribution to
$B-\bar B$, $B_s - \bar B_s$ or $K - \bar K$ mixing, violation of the three
generation CKM unitarity, or a new contribution to decay amplitudes.  Once
some inconsistency within the \sm\ is found, then 
the pattern it exhibits can perhaps be
used to get some insight about the kind of the new physics responsible for 
it.

The ideas presented here should be, of course, additional to other methods
of looking for new physics \cite{NPref}.
New physics can be found in several other ways:
if the values of $\alpha$ and $\beta$ are outside the \sm\ allowed range; 
if the asymmetry in $B_s$ decay mediated by $b \to c \bar c s$ it
significant; or, if asymmetries that should be the same in the \sm\ are
found to be different \cite{gw}.  Because any discrepancy can be an
indication of physics beyond the \sm, it is important to try to have as many
independent tests as possible.

If some of the above hints for new physics were found, 
the ideas we presented
have to be modified. For example, if
$a_{CP}(B \to \phi K_S) \ne a_{CP}(B \to \psi K_S)$ which would
indicate a new contribution to the $b \to s$ transition \cite{gw,LoSo}, 
we will not be able to determine $\sign(\sin\be)$ by
comparing $a_{CP}(B \to \psi K_S)$ to $a_{CP}(B \to D^+ D^-)$.
The underlying assumption in this analysis is that the former
measures $\sin 2\beta$ to very high accuracy. A new significant contribution
to the $b \to s$ transition would invalidate this assumption.

However, in some situations of new physics,
the methods we discuss can still be useful.  For
example, in models where the only significant new physics effects are
significant contribution to the $B -\bar B$ or $K -\bar K$ mixing amplitude
the unitarity triangle can, in principle, be reconstructed.  
However, the combination of
discrete ambiguities and hadronic uncertainties make it 
impractical \cite{gnw}.  Reduction of the ambiguities, in a manner discussed
here, may help in making this program feasible \cite{gnw}.

In our analysis we always care only about a sign of a specific quantity.
Usually, the sign of a specific quantity 
can be determined more easily than its magnitude.
For example, the determination of
$\cos 2 \gamma$ from $B^\pm \to DK^\pm$ 
decays is experimentally very challenging. 
However, even a measurement with large errors may be sufficient for our
purpose.
Of course, if
no choice is found to be consistent across the set of measurements we
have an immediate indication for non-\sm\ physics.

While the methods we describe work in generic points
of the parameter space, there are some values of the angles where they
will not  work. This is the case where some of the quantities we need to determine 
are (very close to) zero. For example, when
$\al = \pi/4$ we have $\cos 2 \al=0$. 
Then, the ambiguity in the value of $\al$ is only two fold, 
but it cannot be removed by the methods we presented. 
We used the ratio  $\cos 2 \al \sin \al/\cos 2 \al$
to determine $\sign(\sin \al)$. However, when 
$\cos 2 \al \approx 0$ we will not be able to measure this ratio.

From the experimental point of view, 
since many of the channels we have discussed have yet to be reliably
observed it is not clear how feasible the comparisons we discuss
will be. All these studies are certainly at least second generation
$B$ factory work, not feasible until large data samples have been
accumulated. For example,
the determinations of $\sign(\sin \phi)$ using exclusive decays
involve comparisons of measured
asymmetries in two different channels. Determining the sign of a 
difference of 
two measured quantities, each of which will have  significant errors, is 
certainly not going to be easy, and will be harder if the actual values
of the asymmetries are small (e.g. if $|\alpha|$ is close to $\pi/2$).

To conclude: we explain how the determination of $\sign(\cos 2\phi)$ and 
$\sign(\sin \phi)$ (for  $\phi=\al,\be$) fully resolve
the 16 fold ambiguity in the values of $\al$ and $\be$ as can be extracted
from CP asymmetries in $B$ decays. 
The determinations of $\sign(\cos 2 \al)$ and
$\sign(\cos 2 \beta)$  are theoretically clean. 
The determination of $\sign(\sin \al)$ and $\sign(\sin \beta)$, 
however, are plagued with
some theoretical input, which, at present, is not reliable.
The hope is that by the time the measurements 
will be carried out, our theoretical toolkit will be improved and we will
be able to calculate more reliably
the sign of the relevant hadronic effects.
From the experimental side,
none of the methods we described is easy to carry out. 
Hopefully, some of them will turn out to be useful.

\acknowledgements
We thank Gerhard Buchalla, Isi Dunietz, Boris Kayser, 
Yossi Nir, Luis Oliver, Lincoln Wolfenstein and Mihir Worah
for helpful discussions. H.Q. also acknowledges the hospitality of the 
Particle Physics Department of the Weizmann Institute of Science where her 
work on this topic began.

{\tighten

}
\end{document}